\overfullrule = 0pt

\font\bigg=cmbx10 at 17.3 truept    \font\bgg=cmbx10 at 12 truept
\font\twelverm=cmr10 scaled 1200    \font\twelvei=cmmi10 scaled 1200
\font\twelvesy=cmsy10 scaled 1200   \font\twelveex=cmex10 scaled 1200
\font\twelvebf=cmbx10 scaled 1200   \font\twelvesl=cmsl10 scaled 1200
\font\twelvett=cmtt10 scaled 1200   \font\twelveit=cmti10 scaled 1200
\def\twelvepoint{\normalbaselineskip=12.4pt
  \abovedisplayskip 12.4pt plus 3pt minus 9pt
  \belowdisplayskip 12.4pt plus 3pt minus 9pt
  \abovedisplayshortskip 0pt plus 3pt
  \belowdisplayshortskip 7.2pt plus 3pt minus 4pt
  \smallskipamount=3.6pt plus1.2pt minus1.2pt
  \medskipamount=7.2pt plus2.4pt minus2.4pt
  \bigskipamount=14.4pt plus4.8pt minus4.8pt
  \def\rm{\fam0\twelverm}          \def\it{\fam\itfam\twelveit}
  \def\sl{\fam\slfam\twelvesl}     \def\bf{\fam\bffam\twelvebf}
  \def\mit{\fam 1}                 \def\cal{\fam 2}
  \def\tt{\twelvett}
  \textfont0=\twelverm   \scriptfont0=\tenrm   \scriptscriptfont0=\sevenrm
  \textfont1=\twelvei    \scriptfont1=\teni    \scriptscriptfont1=\seveni
  \textfont2=\twelvesy   \scriptfont2=\tensy   \scriptscriptfont2=\sevensy
  \textfont3=\twelveex   \scriptfont3=\twelveex  \scriptscriptfont3=\twelveex
  \textfont\itfam=\twelveit
  \textfont\slfam=\twelvesl
  \textfont\bffam=\twelvebf \scriptfont\bffam=\tenbf
  \scriptscriptfont\bffam=\sevenbf
  \normalbaselines\rm}
\def\IZ{\relax\ifmmode\mathchoice
{\hbox{\cmss Z\kern-.4em Z}}{\hbox{\cmss Z\kern-.4em Z}}
{\lower.9pt\hbox{\cmsss Z\kern-.4em Z}}
{\lower1.2pt\hbox{\cmsss Z\kern-.4em Z}}\else{\cmss Z\kern-.4em Z}\fi}

\def\doublespace{\baselineskip=\normalbaselineskip \multiply\baselineskip by 2}

\newcount\equationnumber
\advance\equationnumber by1
\def\ifundefined#1{\expandafter\ifx\csname#1\endcsname\relax}
\def\docref#1{\ifundefined{#1} {\bf ?.?}\message{#1 not yet defined,}
\else \csname#1\endcsname \fi}
\def\autoeqnum{\def\eqlabel##1{\edef##1{\the\equationnumber}}}
\def\no{\eqno(\the\equationnumber){\global\advance\equationnumber by1}}
\newcount\citationnumber
\advance\citationnumber by1
\def\ifundefined#1{\expandafter\ifx\csname#1\endcsname\relax}
\def\cite#1{\ifundefined{#1} {\bf ?.?}\message{#1 not yet defined,}
\else \csname#1\endcsname \fi}
\def\autocite{\def\citelabel##1{\edef##1{\the\citationnumber}\global\advance\citationnumber by1}}
\def\preprintno#1{\rightline{\rm #1}}

\hsize=6.5truein
\hoffset=.1truein
\vsize=8.9truein
\voffset=.05truein
\parskip=\medskipamount
\twelvepoint           
\doublespace
\autocite
\autoeqnum 

\def\s{\scriptstyle}
\def\ss{\scriptscriptstyle}
\def\cl{\centerline}

\def\o{\over}

\vskip -48 truept

\preprintno{ICN-UNAM-97-09}
{ \hfill June 10, 1997}
\break
\vskip 0.6truein

\cl{\bigg {CODON BIAS AND MUTABILITY IN HIV SEQUENCES}}
\vskip 0.3truein
\cl{\bgg H. Waelbroeck}
\vskip\baselineskip
\cl{\it Instituto de Ciencias Nucleares, UNAM,}
\cl{\it Circuito Exterior, A.Postal 70-543,}
\cl{\it M\'exico D.F. 04510.}
\vskip 1truein

\noindent{\bf Abstract:}\ \ A survey of the patterns of synonymous
codon preferences in the HIV {\it env} gene reveals a relation between
the codon bias and the mutability requirements in different regions in
the protein. At hypervariable regions in $gp120$, one finds a greater
proportion of codons that tend to mutate non-synonymously, but to a target 
that is similar in hydrophobicity and volume. We argue that this strategy 
results from a compromise between the selective pressure placed on the virus 
by the induced immune response, which favours amino acid substitutions
in the complementarity determining regions, and the negative selection 
against missense mutations that violate structural constraints of the 
{\it env} protein. 

\

\noindent {\bf Key words:} Adaptive evolution -- Nucleotide substitution -- 
Genotype-phenotype relation -- Codon bias -- HIV

\vfil \eject

\line{\bgg Introduction \hfil}

 The redundancy of the genetic code, particularly the ``codon bias'' 
phenomenon, has drawn a great deal of attention in past years. 

 According to the Neutral Theory, in a finite breeding pool
certain synonyms grow and others disappear due to random sampling alone 
(Kimura 1983). The substitution of a codon by a synonym in large 
portion of the population thus finds its most simple 
explanation as a consequence of neutral drift. 

 However, as Grantham first showed, the patterns of synonymous codon usage
are manifestly non-random. In his own words,``$m-RNA$ sequences contain other 
information than that necessary for coding proteins''(Grantham 1980). 
An example is the almost complete rejection of $CGN$ codons in the $HIV$ 
virus, which results from the $A$-pressure typical of lentiviridiae 
and $CpG$ suppression in eukaryotic viruses (Oliver {\it et al.} 1996 and refs.
therein; van Hemert and Berkhout 1995): Considering only neutral drift,
the codon degeneracy should be resolved independently for each 
$m-RNA$ site since silent substitutions are uncoordinated events. 
In unicellular organisms such as yeast the effective populations are
large and even relatively small selective differences (with selective
coefficientes of order $O(1/N_e$)) can overcome neutral drift. 
Among the proposals that have been put forward to justify small 
selective differences between synonymous codons, one typically finds the 
arguments referring to $m-RNA$ secondary structure 
and base-pairing (Fitch 1980; Miyata 1980a,b), the bias against pretermination 
codons (Fitch 1980; Modiano {\it et al.} 1981) and translational efficiency 
constraints due to relative abundances of isoaccepting aminoacyl $t-RNA$ 
molecules (Ikemura 1981 ; Sharp 1986). 
In multicellular organisms such as mammals where the effective population
sizes are usually small, selection becomes a minor factor but the codon 
bias is still non-random. In such cases the phenomenon is best explained 
by noting that neutral evoution or other factors can lead to 
global fluctuations in the mononucleotide pool and relative abundances of 
transcription enzymes specific to each nucleotide (Grantham 1980),
or directional mutation pressure (Jermiin {\it et al.} 1996).

 The chief purpose of this paper will be to present a new proposal 
on how selection can break the symmetry between synonyms, which 
appears to play an important role in adaptation. But before 
we turn our attention to this proposal it is well worth recalling 
ideas reported recently in this journal (Huynen, 1996).

 As Huynen stressed, the Neutral 
Theory does not exclude the possibility that neutral evolution 
could ``facilitate adaptive evolution by increasing the number
of phenotypes that can be reached with a point mutation from an 
original phenotype''. Huynen 
showed that neutral drift can ``set the stage'' for adaptive 
evolution by carrying the genotype to a point in the landscape 
from which it can jump to a new phenotype with a single point mutation.

 This mechanism is possible because the genotype-phenotype map 
is highly redundant and ``non-trivial'': many sequences encode the 
same phenotype, yet such synonymous sequences can produce different 
mutant phenotypes following a point mutation. An evolving genetic 
population is constantly testing the neighboring phenotypes
through mutation attempts, most of which are negatively
selected, while at the same time random drift proceeds along the ``neutral
net''. When this random drift reaches a point in genotype space from
which an acceptable mutant can be found in the target space, 
a missense mutation can occur without being negatively selected. 

 Once we recognise that different synonymous sequences mutate 
to different ``targets'', one can turn this around by considering 
the fact that each codon in a sequence must itself have originated 
from some precursor through an earlier mutation. Of course different 
codons have different possible precursors. Statistically speaking, this 
implies that codons with a greater number of possible precursors are more 
likely to be found. 

 The prior mutation which gave rise to a given codon may have been a 
silent mutation or an amino acid substitution. In the case when 
the codon occured as the result of an amino acid substitution there is no
guarantee that the mutation can be reversed: further mutations that have 
accumulated since then at other positions in the sequence may have altered the 
structural conditions that prevailed when the precursor amino acid was 
being used. This irreversibility has been observed particularly in highly 
conserved sequences: For example, in cytochrome {\it c} where only 
about 10 \% of all codons are variable in any one mammalian species, 
the set of variable codons or ``covarions'' is known to change as mutations
are fixed (Fitch 1971). However, in highly variable 
sequences such as the $m-RNA$'s that code for HIV proteins, non-local 
structural constraints are not so important and it is reasonable to 
assume that most codons could potentially revert back to their precursors. 

 Thus, as a first approximation for variable sequences we can 
identify the set of possible precursors with the set of possible 
mutation ``targets''. This gives a new dimension to our previous 
comment that ``codons with a greater number of possible 
precursors are more likely to be found'': Indeed, these very same 
codons with a large number of possible precursors are also those 
which can most easily mutate to a new target. One 
implication of this fact is that genetic sequences are  
better able to resist the potentially destructive effect of mutation,
by using codons which mutate more easily to a synonym or to an amino 
acid with similar properties of hydrophobicity and volume. A second and 
less trivial implication is that an evolving species can better take 
advantage of mutations to reach fruitful goals. In the case of 
a retrovirus for example, such a ``fruitful goal'' might be  to
alter a neutralization epitope which has come to be recognised by the 
immune system. The ability of a virus to generate a high variability 
{\it in vivo} may be an essential part of its survival strategy,
for its long incubation period (Nowak 1992). That this strategy could be 
supported in part by the codon bias is an example of the idea which motivated
this paper: The preference for synonyms with better mutability properties 
favours adaptation. 

 From an information-theoretic viewpoint, the claim is that the
information encoded in the distribution of synonyms prepares the 
genetic sequences for the task of eventually mutating when 
required to by the environment. Our arguments above suggest 
that such a ``preparation'' does not imply a violation of causality, if
the adaptation strategies that will be needed in the future are
similar to those that have functioned in the past.

 The HIV-{\it env} proteins provide us with an excellent test-bed 
for these ideas. Neutralising antibodies are produced predominantly 
to an epitope that overlaps with one of the two hypervariable 
regions in gp120, especially the V3 loop. To escape 
the immune system the virus must generate missense mutations in these
regions; this occurs {\it in vivo} on a time scale of about a day 
due to the poor accuracy of reverse transcription. On the other hand
this same lack of accuracy puts tremendous selective pressure for 
the virus to resist transcription errors
in regions that must be conserved, either for structural reasons or
to mediate essential functions of the virus. For example missense 
mutations in the CD4 binding site or the fusogenic domain in gp41
are mostly rejected. 

 This suggests two important properties that should be noticeable 
in the HIV-{\it env} sequences. 

 First, codons in recognition regions (such as the V3 loop) 
are likely to originate from 
a precursor that coded for a {\it different} amino acid, due to the 
selective pressure to adapt to the immune system. Vice-versa,
in conserved regions the precursor is more likely to be a 
synonym. 

 Second, as we explained above, the reversibility of mutations imples 
that the conserved parts of the sequence will use codons that have a 
higher probability to mutate to a synonym, and vice-versa, the V3 loop
will use codons that can better mutate {\it non-synonymously} to 
another acceptable amino acid.

 In the remainder of the paper we will analyse empirical evidence 
for both of these properties. 

\

\line{\bgg Data and Models \hfil}

 We will begin by illustrating the arguments above with the help of a 
simple ``toy model''. 

\

\noindent {\it A toy model: The Met-Leu system}

 Let us first consider a position in the 
protein that is completely conserved and only accepts the amino acid 
Leucine. There are six codons that code for {\it Leu} 
($CTN$ and $TTPu$), but not all have the same number of silent 
point mutations: $CTPy$ codons have three possible silent mutations 
(at the third base), $CTPu$ have four (first base $C \rightarrow T$ 
transition and third base degeneracy) and $TTPu$ codons have two 
possible silent mutations (first base and third base transitions). 

 If we assume that the nine possible point mutations of each codon
are equiprobable, it follows that the success probability
of a point mutation of $CTPu$ is 3/9, for $CTPy$ it is 4/9 
and for $TTPu$ it is only 2/9. Obviously it would be convenient 
for the virus to use the $CTPy$ codons predominantly, to reduce the 
risk of a fatal transcription error. But a codon doesn't reveal 
its mutation success rate until it has actually mutated, so how can 
the choice of the best predominant codon occur without violating 
causality? Since each of 
the six codons represent the same amino acid there is no selective 
pressure in favour of any particular codon. Rather, the point is that
a codon with a higher probability of successful mutation is also the
target of a greater number of possible precursors. 

 A simple experiment demonstrates this point. Let us assume that 
we begin with a gene pool where the six codons are  represented equally.
Each codon grows with a reproduction rate $r$ and is subject to 
nine possible point mutations, which we will assume to be
equiprobable for the sake of simplicity. Discounting the effect
of mutation, each codon would grow by a factor $r$. If we introduce 
a mutation rate $\mu$ this becomes $r(1-\mu)$, plus the gain term 
from other codons that mutate towards it. Since each $CTPu$ codon 
can mutate to three possible synonyms, likewise it can be reached 
from mutations of any one of these three other codons; therefore 
the proportion of $CTPu$ in the population grows initially by a factor
$r(1-\mu+3\mu / 9)$. $CTPy$ grows by $r(1-\mu+4\mu /9)$, and 
$TTPu$ by $r(1-\mu+2\mu /9)$. Thus, after one step of evolution
the codon $CTPy$ is slightly more abundant than the others. The 
evolution equations for this system are the following.

$$x(t+1) = r(1-\mu)x(t) + {\mu \o 9} r x(t) + {2\mu \o 9} r y(t) $$
$$y(t+1) = r(1-\mu)y(t) + {\mu \o 9} r y(t) + {2\mu \o 9} r x(t) 
+ {\mu \o 9} r z(t) $$
$$z(t+1) =  r(1-\mu)z(t) + {\mu \o 9} r z(t) + {\mu \o 9} r y(t) ,$$
where $x(t)$ denotes the portion of the population with the codon $CTPu$
at generation $t$, $y(t)$ is the portion with $CTPy$ and $z(t)$ is the 
portion of the population with the codon $TTPu$.
An integration of this system with $\mu = 0.01$ and $r = 1+2\mu/3$
is shown in [Figure 1]. The most ``mutable'' codon, $CTPy$, reaches 
the highest asymptotic proportion in the population with $2.2\times$ more 
examples than $TTPu$, which is least resistant to mutations.

 The story becomes even more interesting if we consider the interaction 
with the immune system. Again, we construct a simple toy model to 
illustrate our point: We will assume that only two amino acids are 
possible at a given position: {\it Leu} and {\it Met}. 

 With the immune system in mind, one might assume that a codon has 
an extra selective advantage when it is the result of a missense 
mutation in the recent past. This effect can be modeled most 
simply by assuming that {\it only} missense mutations survive. This 
amounts to defining one time step as the time required for the 
induced immune response to eliminate (almost) entirely a detected 
pathogen, or about two weeks.

 Of the six codons that code for {\it Leu} ($CTN$ and $TTPu$), 
only $CTG$ and $TTG$ can mutate to $ATG$ which codes for Methionine. 
Starting once again with all {\it Leu} codons equally represented, after one
time step all codons which are not able to mutate are detected 
by the immune system and destroyed; the population is thus reduced 
to $ATG$, which is generated as a mutant from $CTG$ or $TTG$. At future 
generations we will find this that  {\it Leu} is always coded as 
$CTG$ or $TTG$, which originates from mutations of the {\it Met} codon. Notice 
that these codons are winners because they are mutation {\it targets},  
not because they have better adaptability properties. Yet, obviously,
since mutations are reversible it also happens that these same codons 
are precisely those which have the ability to mutate non-synonymously, 
thereby escaping detection by the immune system. 

 This example shows how adaptation can benefit from the apparently 
trivial fact that the sequences that are most likely to be reached
from a precursor are also those which are best able to mutate. 

 Of course these examples are extremely simplified and are only intended 
to illustrate our point; a more serious  ``genetic algorithm'' model was 
constructed to take into account the finite size effects, including 
neutral drift, and a more realistic evaluation of the selective advantage 
of non-synonymous mutations at regognition sites. The description of this 
model and its results lies beyond the scope of this paper, but will be 
published elsewhere (Mora and Waelbroeck 1997). Suffice it to say in these 
lines that the selection of codons with better mutability properties 
was confirmed with the genetic algorithm model, in spite of the 
fact that the fitness function was strictly symmetric with respect 
to different codons that code for the same amino acid.

\

\noindent {\it Nucleotide sequences}

 We have based our study on the aligned nucleotide sequences from 286 
examples of the HIV {\it env} gene, compiled by Myers in 1994. Some of the 
aligned sequences in this database are substantially shorter or
include large gaps. This indicates that, besides the point 
mutations that we are considering here, the evolution of those
sequences involved also substantial deletion events. Large deletions 
are likely to affect important structural aspects of the protein, 
and this in turn can be expected to affect the issue of which 
mutations are allowed, and vice-versa, which are negatively selected. In 
particular, it may ``freeze in'' some of the amino acid substitutions 
that have occured in the past, blocking the possibility of a reverse 
mutation that would take a codon back to its precursor. Since 
our argument to the effect that evolution favours adaptation 
is based on the idea that point mutations of codons can in principle
be reverted, the effect we are looking for will be clearest if we limit  
to analyzing those sequences that are most similar in length to the consensus. 
For this reason we have discarded the short sequences 
from the set. We also found 9 other sequences with one or two 
premature termination codons; these could be either errors in the 
dataset or real mutations with the same effect as the
previously-discussed large deletions; in both cases the safer 
approach was once again to remove the sequences from the dataset.  

 Of the 286 sequences, 61 involved large gaps and were deleted and
as mentioned 9 more contained premature termination codons, leaving 216 
complete sequences. All remaining sequences still included
some gaps relative to the alignement of 3189 nucleotide sites, but 
were retained since small gaps are less likely 
to alter substantially the structure of the protein or make most 
prior mutations irrreversible. The existence of gaps implies that
the number of codons observed at any given position 
varies, up to a maximum of 216. 

 The overall codon usage patterns in these sequences is represented 
in Table 1. The $A$-pressure is evident, as well as the usual bias 
against $CpG$ codons in lentiviridiae. 

 We have also represented the number of different amino acids observed 
per position, averaged by segments of 20 positions, as a measure of
the variability different segments of the {\it env} protein. One 
can note clearly the conserved regions corresponding to the CD-4 
binding site and the fusogenic domain, which mediate 
essential functions of the virus: The variability in both cases is
extremely low (positions 560-580 and 680-780, Figure 2). Conversely, the 
V3 loop region (positions 380-540) stands out as one of two
hypervariable regions, together with the segment 140-280 in gp120. 

 Another measure of variability is that suggested by the work of 
Almagro, Lara Ochoa and Vargas-Madrazo (Almagro {\it et al.} 1995). 
They showed that key positions in the complementarity determining regions
that are responsible for maintaining canonical structures are characterised 
by an inverse power law distribution of amino acids. At other positions 
an exponential distribution is observed, suggesting that 
negative selection is at work against acids with different 
properties of hydrophobicity and volume. The number of positions in each 
20-position interval where we could identify a power law 
distribution of amino acids, was represented in Figure 3. In this case
the structural regions identified previously are even more clearly 
distinguished, as the number of power law positions in those regions 
is precisely zero.

\

\noindent {\it Probability of origin and ``mutability''}

 We wish to analyse which codons are observed at each position,
and for each codon, where it came from (probability of origin) and
how can it mutate (probability of target). 

\noindent $\bullet$ {\it probability of origin}. Let $n_i$ denote 
the frequency of ocurrence of each codon at a given position, 
$i = 1, \cdots, 64$. We will need the ``mutability matrix'' 
${\cal C}_{ij}$ whose entries are equal to one if there is a 
point mutation which takes codon $i$ to codon $j$, and zero otherwise. 
With no other information at our disposal we 
will assume that the precursor to each codon ($i$) was one of the codons 
that is still observed at that position in one of the sequences in the 
dataset. In other words, a codon $j \neq i$ is a cantidate to have been the 
precursor to $i$ if $n_j > 0$ and ${\cal C}_{ij} = 1$. If there is more than one 
candidate, the probability of each possible precursor is taken to be
proportional to $n_j$, for each $j$ amongst the candidates. The probability
that the codon $i$ originated from a particular {\it amino acid} is the 
sum of the precursor probabilities over all $j$'s which code for
this amino acid. We will call this the {\it probability of origin}. For example,
for the codon $ATG$ ({\it Met}) we have 

$$P(origin(ATG) = Ile) \equiv {n_{\ss ATA} \o n_{\s tot}} + {n_{\ss ATC} \o 
n_{\s tot}} + {n_{\ss ATT} \o n_{\s tot}},$$
where $n_{\s tot}$ denotes the sum of the frequencies of ocurrence of 
all 9 codons which can mutate to $ATG$ with a single point mutation.
In general,

$$P(origin(i) = X) \equiv \sum_{j \in X} {\cal C}_{ij} {n_j \o n_{\s tot}},$$
$$n_{\s tot} = \sum_j {\cal C}_{ij} n_j,$$
where we used the notation $j \in X$ for the codons $j$ that 
code for the amino acid $X$. The probability of synonymous origin is 
represented in [Figure 4] as a function of the position along the sequence.

 To analyse the possible mutation targets of a given codon 
again we will assume that only single point mutations occur (which
is reasonable in view of the absence of double changes in {\it in 
vitro} studies of point mutations (Fitch 1971); there
are nine possible point mutations of any codon, some of which 
will obviously violate contraints on vital functions of the protein,
as for example for any mutation to one of the termination codons. When
the target is not a termination codon it codes for an amino acid,
which may or may not be the same as before the mutation. The 
probability that a particular amino acid is the mutation target of 
a given codon is the number of codons for this amino acid that can
be reached from the original codon, divided by 9. 

$$P(target(i) = X) \equiv \sum_{j \in X} {{\cal C}_{ij} \o 9}.$$

 In the following sections we will also need a measure of 
{\it distance} to quantify the difference between two amino acids. To this 
end we will use the Euclidean distance in hydrophobicity and volume
(Miyata 1979). This is one of the more rudimentary measures and 
surely not the top of the line from the point of view of secondary 
structure analysis; however, it has the virtue of being simple and 
producing a clear categorization of the 20 amino acids into six 
groups. 

 Combining the definitions of the probability of origin and probability
of target with Miyata's distance, we will consider histograms which
represent the probability of origin (target) as a function of distance.
The probability of target histogram for a random sequence of equiprobable 
nucleotides is given in [Figure 5a], as a meter stick to compare results 
later on. The first bin consists in
the average probability that the mutation target is a synonym. The
second bin is the probability that the target is a {\it different} amino
acid with a distance between 0 and 1. The third bin 
corresponds to distances between 1 and 2, the fourth considers targets
with distances between 2 and 3, and so on. The sum of all bins does not
quite add up to one because we are not representing the cases when the 
target is a termination codon. When the 
synonymous codons are taken to be equiprobable but the amino acid
distribution is the same as for the {\it env} protein, one 
obtains [Figure 5b]. It is worth noting that the amino acid usage 
favours non-silent mutations to similar target amino acids over silent
mutations (first three bars). 

\

\line{\bgg Results \hfil}

 The probability of a synonymous origin was computed for each
observed codon at every position. When the precursor is not known
the probability of origin cannot be determined; these cases are 
not considered in the statistics to be presented below. When the 
probability of origin could be determined, we
considered in particular the probability that the precursor was 
a synonym; we will refer to this below as the {\it probability of
synonymous origin}. This quantity was averaged over segments of 60 
nucleotide sites. The results are represented in [Figure 6] (upper
line). Several facts concerning the structure of the protein can
be clearly noted.

\noindent $\bullet$ The first important hypervariable region, 
from positions number 140 to 280, is characterised by a low 
probability of synonymous origin, as is the V3 loop region 
(380-560). 

\noindent $\bullet$ These two hypervariable regions are separated 
by a region in gp120 which is characterised by both a low 
variability and a high probability of synonymous origin. Likewise
for the part of gp120 nearest to the $NH_2$ termination (positions
1-140).

\noindent $\bullet$ Sharp peaks at positions 560-580 and 
680-780 imply a high probability of synonymous origin in the 
conserved regions corresponding to the CD4 binding site and 
the fusogenic domain and surrounding region.

\noindent $\bullet$ In the remainder of the sequence, one notes an 
alternance of variable regions with conserved regions, with a
clear preference for a synonymous origin in conserved regions, and
vice versa, a low probability of synonymous origin in the 
variable regions. One notes for example the short hypervariable region 
in gp41, pos. 800-820, and the transmembrane segment (860-890) where 
hydrophobic amino acids are dominant. 
 
 Next we consider the probability that the {\it target} following a 
point mutation is a synonym. In this analysis the nine possible 
point mutations of each codon are assumed to be equiprobable. 
The probability of a synonymous target is averaged 
over all codons observed in a segment of 60 nucleotide sites, 
whether or not its probability of origin is known. In spite of the
different averaging sets the probability of a synonymous target 
(lower line in [Figure 6] is clearly correlated to the probability 
of a synonymous origin (upper line in [Figure 6]. In [Figure 7]
we represented the two probabilities for each codon with a 
well-defined probability of origin, with no averaging whatsoever.
Again the correlation is evident: on average, codons which originate
from a synonym tend to have a higher probability to mutate to
a synonym. Two horizontal lines can be clearly noted. One 
corresponds to the amino acids with complete third base degeneracy 
which have a probability of synonymous target equal to 1/3 regardless
of the chosen codon. The other line, at the level 1/9, corresponds 
to amino acids which can be represented by two codons, i.e. where 
only third-base transitions are silent. 

 These results support our claim that the codons that are most 
likely to arise from a prior silent mutation are also those most 
likely to mutate to a synonym. Vice-versa, those that are more likely to 
arise from a missense mutation (e.g. due to the need to adapt to
the immune system) are most likely to again mutate non-synonymously.
This is true because point mutations are mostly reversible. Thus, 
regions that are required to be conserved develop a resistance 
to the high error rate of reverse transcription, while on the 
contrary segments which code for potential neutralizing epitopes 
develop an enhanced ability to generate non-synonymous mutations 
that allow the virus {\it in vivo} to escape detection by the immune 
system.

 The probability of origin and probability of target are best 
represented in bar graphs, where one averages over regions of the 
{\it env} protein previously identified with respect to their 
function, but in exchange show more detailed information about 
{\it which} origins or targets are involved. Once again, the
basr represent distance intervals of length one,
where we are using Miyata's Euclidean distance in hydrophobicity and 
volume. The first bar corresponds to silent mutations only.

 For the first interval (positions 1-140), corresponding mostly to a 
conserved region near the $NH_2$ termination of gp120, the bar 
graphs [Figure 8a,b] are essentially undistinguishable from the average over the 
entire protein which we analyzed in the previous section [Figure 5b]. 
The probability of origin differs from the target distribution 
mostly by the effect of negative selection against missense 
mutations at the conserved sites: since some of the non-synonymous 
targets are rejected, the probability of a synonymous origin is 
greater than the probability of a synonymous target. The negative 
selection is most pronounced for distances greater than 2 (bars 
number 3 and up), indicating once again the importance of structural
constraints. Similar results are obtained for all other intervals 
except the two hypervariable regions in gp120.

 The two hypervariable regions which are known to be involved in 
the recognition of the virus by T-cell receptors (particularly the V3 loop 
region), present a very different situation. The bar graphs in both 
regions are very similar, so we will describe only the first hypervariable 
region of gp120 as an example [Figure 9a,b]. The probability of a synonymous
origin is equal to 22\%, precisely the same as the probability of a 
synonymous target! This leaves only two possible hypotheses: 
either there is no negative 
selection against missense mutations whatsoever, which seems hard to
believe, or the negative selection at a relatively limited number 
of sites is compensated by a contrary {\it positive} selection of 
non-synonymous mutations that help the virus escape the induced immune
response, at sites that are involved in one way or another 
with a neutralization epitope. Looking at the complete histograms 
reveals that this second hypothesis is the correct one. The mutation
targets at distance greater than zero but less than two amount to 
27\% + 18 \% = 45\% of the total; amino acids at distance 2 or 
greater are targeted 33\% of the times. The probability of origin is
notably different: 52\% of the precursor codons code for an 
amino acid with similar hydrophobicity and volume properties, while 
only 26\% originate from amino acids with distances greater than 2.
This shows that negative selection is at work against targets which 
alter the amino acids' properties substantially, since 33\% of 
the targets will have this property but only 26\% are fixed. On 
the other hand, and even more strikingly, the
probability that a codon originates from a missense mutation with
distance less than two is {\it greater} than the corresponding 
target probability: this indicates that selection favours the most 
those mutations that alter the amino acid, but not so much as to 
violate structural constraints.

 This shows that positive selection in favour of non-synonymous
mutations can be observed through the patterns of codon 
usage. This claim is also supported by comparing the 
probability of synonymous origin for the different hypervariable 
regions which we indentified in the previous section [Figure 2]. 
The probability of synonymous origin in
variable regions {\it not} related to neutralization epitopes, 
such as 800-820 in gp40, is not nearly as low as for the hypervariable 
regions in gp120. This indicates that in these variable regions 
one has weak structural constraints but not the positive selection
in favour of missense mutations at potential neutralization 
epitopes.

\

\noindent {\it Pretermination codons and the Modiano hypothesis}

 It is well worth recalling in this context Modiano's proposal 
that codons capable of mutating to termination codons are disfavoured
as an evolutionary strategy, to enhance the resistance to mutations 
(Modiano {\it et al.} 1981). 
This proposal is related to ours, although it is clearly less 
ambitious in its scope in that it refers only to one particular 
example of how the choice between synonymous codons can help to 
improve the mutability properties of a gene. Modiano's proposal
stems from the observation (Fitch 1971) that such ``pretermination codons'' 
are significantly less frequent than their synonymous partners. 
However the proposal has been sharply criticised for several reasons.
Kimura, insiting that this effect is the result of neutral drift
and aminoacyl $t-RNA$ relative abundances, has pointed out that the 
selective advantage of a bias against 
pretermination codons in species with a very low mutation rate 
would be extremely small, making it unlikely to impose 
itself against the neutral drift phenomenon: Indeed,  
random sampling dominates over selection when the effective 
selective coefficient of a genetic property is smaller than $1/N_e$, where 
$N_e$ is the effective breeding population (Kimura 1983). Moreover, in the
absence of a specific mechanism to explain how the selective 
disadvantage of a pretermination codon can manifest itself 
one falls into the trap of causality violation: How can a codon 
``know'' that it is a pretermination codon without actually 
falling victim to this fatal mutation, and once it has fallen 
victim to the mutation how can the genetic system recover the 
information that this was a bad choice of codon which should be 
avoided in the future? 

 Our arguments in this paper suggest a possible answer to both criticisms.
Assuming for the argument's sake that we are considering a position
where the {\it only} selective constraint is the negative selection against 
termination codons, the number of possible precursors of any codon
at that position would be equal to 9, except for the pretermination 
codons ({\it TAG, TAA, TGA}). On the other hand, pretermination 
codons can be reached by point mutation from a smaller number 
of possible origins: for example, {\it TAT} ({\it Tyr}) has only 7
possible precursors. So there is a mechanism whereby pretermination 
codons can be disfavoured without violating causality, furthermore 
this mechanism is independent of the mutation rate since a codon
necessarily must have originated from some alloed precursor regardless
of how long ago the mutation is expected to have occured.

 Yet it is clear that this argument is insufficient to explain the
observed effect since it predicts a negative bias against 
pretermination codons of only -2/9 for {\it TAC, TAT, TCA, TTA, TGG} 
and -1/9 for {\it TGC, TGT, TCG, TTG, CAA, CGA, CAG, AAA, AAG, AGA, 
GAA, GAG, GGA}. So the much stronger bias noted originally by Fitch 
remains mostly unexplained. 

 The case of the AIDS virus is different from that considered by 
Fitch, and due to its high mutation rate it can be assumed to 
be a completely independent neutral drift experiment, so it is
well worth analyzing once again the evidence for a bias against 
pretermination codons with this particular dataset. In a random 
sequence of equiprobable nucleotides, the expected number of point 
mutations from an expressed codon to a termination codon would be 
0.4219. Choosing codons at random but respecting the observed frequency 
of amino acids in the {\it env} protein, this number drops to 0.4058; an 
effect which can only be attributed to the selection of amino 
acids. For the actual codons from the {\it env} sequences, the average 
number of point mutations to termination codons is 0.4571! This 
indicates that other factors besides the risk of mutating to a 
termination codon are more significant. This should come as no surprise,
since we have just shown evidence in favour of two such competing factors:
The need to avoid missense mutations at conserved sites, and the 
positive selection of amino acid substitutions with distances less than 
two in the regions involved in the recognition mechanism. 

 Taking once again the {\it env} protein data by segments 
of 60 nucleotide sites, we computed the average at each position 
of the number of possible point mutations to a termination codon
[Figure 10]. For the sake of comparison, we also 
computed the expected number of mutations to a synonym; with the 
exception of the recognition regions one can assume that such silent 
mutations are always fully accepted, contrary to mutations
to a termination codon [Figure 11] (this differs from the 
probability of synonymous origin in the averaging method: here,
we are considering a partition of the identity at each position, 
whilst in [Figure 6] we averaged over all codons which had a 
well-defined probability of origin regardless of how many times
it was observed in the dataset). There are strong fluctuations of 
the probability to mutate to a termination codon, which do not appear
to be correlated to the structure of the {\it env} protein.

 The conclusion is that in considering arguments that refer to the 
mutability properties of a codon one must look at all nine possible 
mutations before deciding whether a particular codon should be 
considered to be ``better'' than other synonyms. The codon that has 
the best chance of mutating to another allowed amino acid and fool 
the immune system may well be that which most risks mutating to a 
termination codon. 
\

\line{\bgg Discussion \hfil}

 It is well worth recalling at this point recall Huynen's 
argument, that neutral evolution 
could ``facilitate adaptive evolution by increasing the number
of phenotypes that can be reached with a point mutation from an 
original phenotype'' (Huynen, 1996). His claim 
was that random drift eventually takes the genetic sequences to
a point where a jump to a new phenotype can occur with a high
probability. Our result, in short, is that random drift is not 
alone to carry out this task: selection forces also drive the system
towards regions of the sequence space that have better ``targets''
following point mutations, thereby setting the stage for both 
a better degree of resistance to mutation and improved adaptation
capabilities.

 In our analysis we always considered that the nine possible 
point mutations of each codon are equiprobable, i.e. that there is 
no directional mutation pressure or bias of any kind. The reason 
for this attitude is that we wanted to prove that information 
about phenotypic mutations could be introduced {\it via} the choice of 
symmetry breaking alone, {\it without} the need to postulate 
directionality in the mutations themselves: Such a directionality in 
fact {\it can} exist and may play a selective function for example 
in avoiding pretermination codons. However, more complex phenotypic 
mutations cannot be prepared in this way without violating 
the ``central dogma'': such mutations would require that directional 
mutation pressure be applied differentially to different segments of 
chromosome and not globally to the entire transcription process; the 
only way that this could only happen is if proteins could
carry information about the environment into the transcription
machinery. This applies to all global selective factors or 
other sources of directional mutation. A great variety of mechanisms
have been proposed, most notably the selective effect of differential 
aminoacyl $t-RNA$ availabilities, $m-RNA$ secondary structure constraints, 
fluctuations in mononucleotide transcription enzyme availabilities that induces 
directional mutational pressure in viral $RNA$ transcription: all of these
mechanisms are likely to be important in understanding the codon
bias, but none of them can help to explain how  
evolution at phenotypic level can be organised to better meet the
varying demands of the environment.

 The mechanism we have proposed in this paper, on the other hand, 
can in principle allow genetic systems to organise phenotypic mutations 
without violating the central dogma, since no information is introduced 
at the level of mutations at the genotype.   

 To understand how this mechanism can be promoted from the level 
of codon bias, where it is very weak and not all that relevant 
as we saw in the results above, to the level of organizing 
complex phenotypic traits, requires one to consider the non-trivial
computation carried out by the biochemical processes involved in 
interpreting the genetic information stored in the genotype, to define
the shape and function of macroscopic phenotypic traits.  

 The concept of synonym must be generalised beyond the codon-aminoacid
redundancy, for this mechanism to be useful towards understanding the 
self-organization of phenotypic evolution in complex organisms. The
chromosome does not encode directly the size and shape of various parts of
an organism, but instead an {\it interpreter}, embodied by biochemical 
processes in living cells (and amongst them), translates the genotype 
into a phenotype. In this translation there are many possible sources 
of redundancy, the codon bias being only a relatively insignificant 
example. There are more subtle forms of synonyms, involving issues 
from protein secondary structure to the machinery of gene regulation, 
for which symmetry breaking can be related to the emergence of an {\it 
algorithmic language.}

 Considering the chromosome (genotype) as an algorithm, the interpreter is
the ''computer'' which executes the algorithm and the phenotype is the
solution. In this sense, the breaking of symmetry is related to the
selection of a language, where ''words'' or ''grammatical rules'' are
selected in order to facilitate the search of well-adapted offspring, ({\it %
i.e.} successful mutants). The identification of such subunits of 
genetic information (Schmitt, 1996) to facilitate the search for mutant phenotypes is
related to the standard A.I. problem of finding an approximate decomposition
of an optimization problem into smaller subproblems. The condition for 
such a strategy to succeed is that when the solutions to the subproblems 
are reached then a good approximation to the global solution is reached
as well. This requires that the fitness landscape should have a certain 
amount of structure; by unravelling this structure the emergent language 
results in an effective smoothing the induced landscape on genotypes. 
By ``effective smoothing'' we mean the population-dependent property that 
mutations at the level of the genotype have better mutation targets on 
average than in a random population. This implies first of all a solution 
of the brittleness problem, since the first task is that mutant algorithms
be meaningful, and secondly an enhanced ability to produce genetic 
improvements (better algorithms). For the proposed mechanism to work it
is necessary that the landscape be sufficiently correlated, and 
that the interpreter be well adjusted to the structure of the problem. An
example would be the Kauffman's {\it Nk} landscapes for $k<N$, 
together with his model of cellular automata for gene regulation. Another
example (Angeles 1997) is the cell division interpreter in Kitano's
neurogenetic model (Kitano 1990, 1994).

 In order that the symmetry breaking which necessarily relfects only 
{\it past} adaptation pressures should favour the search of {\it future} 
solutions, the evolution of the landscape must respect certain rules. 
Namely, the decomposition of the optimization problem into subproblems
must be independent of time, so that the algorithmic language which has 
been successful in past should continue to be useful in future.
This is the requirement of {\it structural decomposition stability}: 
The landscape evolution must preserve the structural decomposition 
of the adaptation problem.

 One might conjecture that extinctions are related to a violation of
structural decomposition stability. For instance, the algorithmic language
guiding the search of new dinosaur species would presumably have been
incapable of producing viable solutions in the environment which is assumed
to have provoke their demise.

 The symmetry breaking which we observe in these experiments support these 
ideas by suggesting that with a less trivial interpreter one might witness 
the emergence of an ``algorithmic language'' tuned to the interpreter. 

 We are currently analyzing several Genetic Algorithm models to this effect,
(Holland, 1975; Goldberg, 1989),
using certain classes of controlled rugged landscapes that are more
realistic from a biological point of view (Kauffman, 1989,1990,1993).
A related challenge is to exploit the emergence of a language to assist 
in the design of a new generation of genetic algorithms as an improved 
general purpose optimisation method. A key for success in this direction is 
the codification method: The interpreter should have the sufficient 
flexibility to be able to solve the decomposition problem, but not so much
flexibility that it could solve any possible problem, since in that case the
search space for the desired algorithmic language would be far too large. 
Another application of the language emergence is the development of an GA 
to perform complex computational tasks (Crutchfield 1994), such as combinatorics. 
Interesting applications may also follow in adaptive systems modelling 
where adaptability is an important property, for example
the forecasting problem in financial markets. 

\

\noindent {\bf Acknowledgements} Most of the numerical results presented
in this work were derived with a code which originated from the work of 
Sara Vera Noguez. We thank Juan Carlos Almagro, Germinal Cocho, Gustavo 
Mart\'\i nez-Mekler, Chris Stephens and Magali Waelbroeck for many enlightening 
discussions, as well as to the entire Complex Systems group under 
{\it NNCP} (http://luthien.nuclecu.unam.mx/~nncp), for maintaining a 
stimulating research atmosphere. This work is supported in part by the 
DGAPA-UNAM. 

\vfil\eject

\line{\bgg REFERENCES \hfil}

\noindent Almagro J C, Vargas-Madrazo E and Lara-Ochoa F (1995) Distributions of the 
Use Frequencies of Amino Acids in the Hypervariable Regions of Immunoglobulins.
J Mol Evol 41: 98-103

\noindent Angeles O Stephens C and Waelbroeck H (1997) Emergence of algorithmic 
languages in genetic systems. National University of Mexico Preprint 
ICN-UNAM-97-08 (adap-org/97-08-001)

\noindent Crutchfield J P (1994) The Calculi of Emergence: Computation, Dynamics
and Induction. Physica {\bf D 75}: 11-54

\noindent Fitch W M (1971) Rate of Change of Concomitantly Variable Codons. J Mol
Evol 1: 84-96

\noindent Fitch W M (1980) Estimating the Total Number of Nucleotide Substitutions 
Since the Common Ancestor of a Pair of Homologous Genes: Comparison of Several
Methods and Three Beta Hemoglobin Messenger RNA's. J Mol Evol 16:153-209

\noindent Goldberg D E (1989) Genetic Algorithms in Search, 
Optimization and Machine Learning. Addison Wesley, Reading, MA.

\noindent Grantham R (1980) Workings of the Genetic Code. Trends in Biochemical 
Sciences 5:327-331

\noindent Holland J H (1975) Adaptation in Natural and Artificial Systems.
MIT Press, Cambridge, MA. 

\noindent Huynen M (1996) Exploring Phenotype Space Through Neutral Evolution. J 
Mol Evol 43: 165-169

\noindent Ikemura, T (1981) Correlation between the Abundance of {\it Escherichia coli} 
Transfer RNAs and the Occurrence of the Respective Codons in its Protein 
Genes: A Proposal for a Synonymous Codon Choice that is Optimal for the
{\it E. coli} Translational System. J Mol Evol 151: 389-409

\noindent Jermiin L S {\it et al.} (1996) Unbiased Estimation of Symmetrical Mutation
Pressure from Protein-Coding DNA. J Mol Evol 42: 476-480

\noindent Kauffman S A (1989) Adaptation on rugged fitness landscapes, in: 
Lectures in the Sciences of Complexity. D. Stein (ed.) (Reading, 
MA, Addison-Wesley) pp. 527-618

\noindent Kauffman S A (1990) Dyanamics of Evolution. Lecture at the Workshop 
Complex Dynamics and Biological Evolution. Hindsgavl, Physica {\bf D42}:135

\noindent Kauffman S A (1993) The Origins of Order. Self-Organization and 
selection in Evolution. Oxford University Press, New York, Oxford

\noindent Kimura M (1983) The Neutral Theory of Molecular Evolution. Cambridge University
Press, Cambridge

\noindent Kitano H (1990) Designing neural networks using genetic algorithms with graph 
generation system. Complex Syst. 4: 461-476

\noindent Kitano H (1994) Neurogenetic learning: an integrated method of designing and 
training neural networks using genetic algorithms. Physica D 75: 225-238

\noindent Miyata T, Miyazawa S and Yasunaga T (1979) Two Types of Amino Acid 
Substitutions in Protein Evolution. J Mol Evol 12:219-236

\noindent Miyata T and Yasanuga T (1980) Molecular Evolution of mRNA: A Metho for Estimating 
Evolutionary Rates of Synonymous and Amino Acid Substitutions from 
Homologous Nucleotide Sequences and Its Applications. J Mol Evol 16: 23-36

\noindent Miyata T, Yasanuga T and Nishida T (1980) Nucleotide Sequence Divergence 
and Functional Constraint in {\it mRNA} Evolution. Proc. Natl. Acad. Sci. USA 
77:7328-7332

\noindent Modiano, G, Batistuzzi, G and Motulsky, A G (1981) 
Nonrandom Patterns of Codon Usage and of Nucleotide Substitutions in $\alpha$- 
and $\beta$-globin genes: an Evolutionary Strategy Reducing the Rate of 
Mutations With Drastic Effects? Proc. Natl. Acad. Sci. USA 78:1110-1114

\noindent Mora Vargas J, Stephens C and Waelbroeck H (1997) Symmetry 
Breaking and Adaptation: Evidence From a ``Toy Model'' of a Virus. 
National University of Mexico Preprint ICN-UNAM-97-10 (adap-org/97-07)

\noindent Nowak M A (1992) Variability of HIV Infections. J. theor. Biol. 
{\bf 155}: 1-20

\noindent Oliver, J L and Mar\'\i n, A (1996) A Relationship Between GC Content and 
Coding-Sequence Length. J Mol Evol 43:216-223

\noindent Schmitt A.O. {\it et al} (1996)  The modular structure of informational 
sequences. Biosystems {\bf 37}, pp. 199-210

\noindent van Hemert F J and Berkhout B (1995) The Tendency of Lentiviral Open 
Reading Frames to Become A-Rich: Constraints Imposed by Viral Genome 
Organization and Cellular $tRNA$ Availability. J Mol Evol 41:132-140

\vfil\eject

\centerline{FIGURE CAPTIONS}

\

\noindent {\bf Figure 1} The evolution of the codon abundance for the amino
acid {\it leucine} is represented in a simple model where all six possible 
codons for this amino acid have the same fitness value and the codons are 
subject to single point mutation only. The codons $CTPy$ (upper line) are 
favoured because they are most easily reached from mutations of other 
leucine codons. For the same reason they are also more resistant
to point mutation, as four of the nine possible point mutations produce
other leucine codons. The least mutable codons are $TTPu$, which
allow only two possible silent mutations (lower line). Their relative 
abundance drops to about half that of $CTPy$ codons. This illustrates 
our main point, that synonyms are not equal in molecular evolution, because 
the more ``mutable'' ones have an effective selective advantage in the 
long run due to their ability to generate successful offspring.

\

\noindent {\bf Table 1} The codon usage patterns are shown for the AIDS 
{\it env} data considered in our analysis. The different groups of amino
acids with similar properties of hydrophobicity and volume are grouped 
together. Counterclockwise from the upper left corner, we find the following 
groups: 1 (special), 4B (large hydrophobic), 4A (small hydrophobic), 2 (neutral),
3A (small hydrophilic) and 3B (large hydrophilic).

\

\noindent {\bf Figure 2} The number of amino acids observed per position, 
averaged by intervals of 20 positions, is represented. The two peaks are
both in the $gp120$ segment of the protein, the second one being the V3 loop
region.

\

\noindent {\bf Figure 3} The average number of positions for which the 
distribution of amino acids matches a power law distribution is represented.
The two peaks again correspond to exposed parts of $gp120$, whilst  
structural regions of the {\it env} protein such as the fusogenic domain
(680-780) are clearly identifiable by the absence of power-law distributions. 

\

\noindent {\bf Figure 4} The probability that a codon for an amino acid $A$ 
originated by point mutation from a precursor that coded for an amino acid $B$
is given, as a function of the distance between these two amino acids. In 
the bar graph, the first bar corresponds to distance zero, i.e. it is 
the probability of origin by silent mutation. The next bars are for unitary 
distance intervals starting from the interval (0, 1]. Miyata's Euclidean 
distance in hydrophobicity and volume was used.

\

\noindent {\bf Figure 5} The probability of a target by point mutation 
is represented as a function 
of the Miyata distance between the two amino acids, $d(A,B)$. Again, the first 
bar in these graph correspond to silent mutations while the following bars 
represent unitary distance intervals starting from the interval 
$d(A,B) \in (0,1]$. In Fig. 5a the target probability is given for a 
random sequence of equiprobable nucleotides; in Fig. 5b the actual 
amino acid distribution of the {\it env} protein was used but codons were 
chosen at random among the synonyms that code for each amino acid. Comparing
the first three bars of each graph, one notes that the amino acid usage 
favours non-silent mutations to targets with distances $d(A,B) \leq 2$ over 
neutral mutations.

\

\noindent {\bf Figure 6} The probability of a synonymous origin (solid line)
and probability of synonymous target (dashed line) are represented, averaged
over segments of 20 positions along the {\it env} protein. In the first case,
one looks for possible point-mutation precursors of each codon 
in the dataset and estimates the probability that the precursor was a 
synonym. The probability of a syonymous target is evaluated by assuming 
that the nine possible point mutations of a codon are equiprobable;
for example for third-base degenerate codons the probability of a synonymous
target is 3/9. The solid line is always above the dashed line because silent 
mutations are usually not rejected; the difference between the probability
of a synonymous target and the probability of a synonymous origin indicates the 
degree of negative selection against missense mutations. One notes the 
correlation between the two lines, as well as the two windows with low 
probabilities of synonymous origin/target and weak negative selection
in gp120 (positions 140-280 and 380-560).

\

\noindent {\bf Figure 7} The probability of synonymous target is plotted as 
a function of the probability of a synonymous origin. The correlation can 
be noted as an upward tilt of this cloud of points. The horizontal lines
at synonymous target probability 1/9 and 3/9 are clearly related to amino
acids with partial and full third base degeneracy.

\

\noindent {\bf Figure 8} The probabilities of origin (Fig. 8a) and target 
(Fig. 8b) are represented as a function of the Miyata distance between 
two amino acids, in the segment 1-140 from gp120 which does not interact 
directly with immunoglobulin complementarity determining regions. The 
negative selection against non-synonymous targets with distances greater 
than 2 is evident: 33\% of all targets are amino acids with a distance 
greater than two but only 21\% of the precursors of observed codons have 
that property (bars 4-7). Vice-versa, only 20 \% of all targets are synonymous 
but only 33\% of codons originate from a silent mutation (first bars).

\

\noindent {\bf Figure 9} Similar graphs as for Figure 8 are given but 
for positions 140-280, an exposed hypervariable region of gp120. Here 
the probability that a codon originates from a non-synonymous precursor 
with Miyata distance between 0 and 2 is {\it greater} than the 
corresponding target probability. This is evidence 
for positive selection in favour of non-silent mutations which help the virus
escape detection by the immune system.

\

\noindent {\bf Figure 10} The average number of point mutations that would 
yield a termination codon is represented, as a function of the position along 
the sequence. There is no evidence for a bias against pretermination codons,
nor any obvious relation between the shape of this curve and the structure of 
the protein. This indicates that Modiano's proposed bias against pretermination
codons is overruled by other factors, such as the bias in favour  
mutations to synonyms, or, in the recognition regions, to different amino 
acids with similar properties. 

\

\noindent {\bf Figure 11} The average number of point mutations that would 
yield a synonymous codon is represented, as a function of the position along 
the sequence. The bias against ``presynonymous codons'' is particularly 
noteworthy in the ranges 240-260 and 500-520 in gp120, both of which are 
potential recognition sites. 

\end